**Catalyzing Equity in STEM Teams: Harnessing Generative AI for Inclusion and Diversity**

Nia Nixon, Yiwen Lin & Lauren Snow

University of California, Irvine

**Author Note**
Nia M. M. Nixon, University of California, Irvine EDUC 3361A, Irvine, California, 92697.
Email dowelln@uci.edu
Yiwen Lin, University of California, Irvine
Lauren Snow, University of California, Irvine

**Acknowledgments:** This work was partially supported by the National Science Foundation (Grant Number 1535300), and National Institutes of Health (Grant Number 5UC2NS128361-02).

*Correspondence: Nia Nixon, School of Education and Information, University of California, Irvine EDUC 3361A, Irvine, California. Phone: 901-409-7171, Email: dowelln@uci.edu.

## Abstract

Collaboration is key to STEM, where multidisciplinary team research can solve complex problems. However, inequality in STEM fields hinders their full potential, due to persistent psychological barriers in underrepresented students' experience. This paper documents teamwork in STEM and explores the transformative potential of computational modeling and generative AI in promoting STEM-team diversity and inclusion. Leveraging generative AI, this paper outlines two primary areas for advancing diversity, equity, and inclusion. First, formalizing collaboration assessment with inclusive analytics can capture fine-grained learner behavior. Second, adaptive, personalized AI systems can support diversity and inclusion in STEM teams. Four policy recommendations highlight AI's capacity: formalized collaborative skill assessment, inclusive analytics, funding for socio-cognitive research, human-AI teaming for inclusion training. Researchers, educators, policymakers can build an equitable STEM ecosystem. This roadmap advances AI-enhanced collaboration, offering a vision for the future of STEM where diverse voices are actively encouraged and heard within collaborative scientific endeavors.

Science, technology, engineering, and mathematics (STEM) stand as the vanguard of modern innovation and societal transformation. These disciplines have accelerated innovation and powered breakthroughs that have reshaped the world. Yet, navigating the complex landscape of scientific progress and technological evolution confronts a persistent challenge: the underrepresentation of diverse voices in STEM (Dasgupta & Stout, 2014; National Science Foundation, 2019). Although the benefits of diversity in fostering creativity and innovation are well-established (Salazar et al., 2017; Yang et al, 2022), the echo chambers of homogeneity within STEM fields continue to impede the full realization of their potential.

The power of diversity in fueling STEM innovation cannot be overstated. Diverse teams bring together a rich tapestry of perspectives, experiences, and skills, igniting creativity and enhancing decision-making (Page, 2007). Despite an increased participation of marginalized groups in STEM over the years, progress has been slow and uneven across fields, particularly in computing and engineering (Fry et al., 2021). The walls of underrepresentation could be costly, as they exclude voices that could contribute significantly to the collective scientific endeavor. The longstanding exclusionary norms and values within STEM culture pose challenges to change (Leibnitz et al., 2022). Barriers derived from the intersecting systems of power, privilege, and oppression, as well as those shaped by gender, race, ethnicity, sexuality, disability, nationality, class and more, can exacerbate the gap in achievement and experience for marginalized individuals in STEM (Collins, 2015; Crenshaw, 1989; Leibnitz et al., 2022). Recognizing this challenge, broadening STEM participation has become a paramount concern for educational researchers, policymakers, and all those invested in the future of STEM.

One effective way to increase equality in STEM is by recognizing the underlying psychological needs of minoritized students (Casad et al, 2018); effective interventions target students' identity, motivation, and belonging (Walton & Cohen, 2011), to name a few. For instance, some interventions facilitate students' appreciating the utilitarian value of their subject (Asher et al., 2023). Others encourage a sense of belonging through gender-paired mentorship (Moghe et al., 2021)—effective in increasing women's retention in STEM. Equal gender groups as a mechanism to mitigate the experience of minoritized students in hostile environments. This intermediate environment creates a shield from stereotype threats and enhances female learners' motivation and STEM career aspiration (Dasgupta et al., 2015).

Artificial intelligence (AI) has been transforming STEM education, creating unprecedented opportunities to address the challenges of diversity and inclusion (Roscoe et al, 2022). In the past twenty years, AI's substantial fertilization of STEM learning prediction, content delivery, assessment, behavior detection, automation, as well as personalization and adaptive learning through Intelligent Tutoring Systems (Baker, 2016; Nkambou et al., 2010). As we move forward, AI will not only continue to expand its capacity in the aforementioned areas but should also be positioned to play a more prominent role in learner-based activities (e.g., project-based learning and collaborative problem-solving) rather than instructor-centered paradigms (Xu & Ouyang, 2022). The emergence of human-AI teaming, for instance, raises promises of AI providing personalized resources and cost-effective solutions for equitable educational outcomes and inclusivity (Chine et al., 2022). A burgeoning amount of research is accelerating to understand how to best utilize Generative AI (GenAI), such as ChatGPT, for educational purposes. As we are still in the early days of GenAI, a more nuanced understanding of the implications of AI integration in academic settings and more empirical studies to examine its effect are needed. It is also imperative to expand discussions amongst educators, technology developers, and policymakers around how AI can contribute to broadening STEM participation and equity for future education.

Our goal is to provide a roadmap for harnessing the power of generative AI for natural language to promote diversity and inclusion in STEM education, particularly through the lens of collaborative problem-solving in STEM. In the following sections, we first outline psychological barriers often attributed to the underrepresentation of women and racial minorities. Second, we discuss opportunities of leveraging AI to foster diversity and inclusion in STEM teams. We follow with an outline of four key policy recommendations which center around the formal assessment of collaborative skills, inclusive analytics in teams, funding socio-cognitive and emotion research, and human-AI teaming for inclusive collaboration. Lastly, we highlight the importance of strengthening stakeholder collaboration amongst educators, students, policymakers, and industry partners. Our aim is to provide one path toward a more diverse, inclusive, and equitable STEM education ecosystem, where generative AI plays a central role in shaping a future where every voice is not only heard but actively encouraged to contribute to the scientific tapestry.

**Psychological Barriers for Underrepresented STEM Students**

In STEM settings that lack gender and racial diversity, individuals from minoritized groups often find themselves to be a rare representative of their group. Skewed demographic representation in these contexts automatically activates psychological stereotypes, raising doubt about the minority's ability in that context, promoting stereotype-driven assumptions about their likely performance (Chatman & Spataro, 2015). For example, many STEM environments are subtly unfriendly or sometimes overtly hostile to women (Logel et al., 2009; Settles et al., 2006).

The numeric scarcity of women in STEM contexts (Dasgupta et al., 2015; National Science Foundation, 2019), nonverbal behavior from male colleagues that excludes women from professional conversations, the use of masculine pronouns to refer to all scientists and engineers, and the prevalence of sexist jokes all signal to women that they are outsiders in STEM, eroding their belonging and self-efficacy, leading to burnout and attrition from STEM (Hall et al., 2015). For minoritized students underrepresented in higher education (e.g., African Americans, Latinx, and Native Americans), stereotypes impugning their group's intelligence and academic ability loom large (Ashburn-Nardo & Johnson, 2008). Although all college students have periodic concerns about their performance, racial minority students who are rare in academic contexts, not only risk failure and embarrassment, but also risk confirming negative stereotypes if they perform poorly. This additional stress can impair their academic performance, through processes involving stereotype threat (Ployhart et al., 2003; Steele & Aronson, 1995).

Teamwork plays a pivotal role in STEM education and professional practice, so these settings are critical for examining psychological barriers. Focusing specifically on the demographic composition of small teams—examining their impact on minority and majority team members— team demographic composition negatively impacts minority participation, motivation, anxiety, self-efficacy, and team satisfaction and performance (Paletz et al., 2004; Pollak & Niemann, 1998; Sekaquaptewa et al., 2007; Stangor et al., 1998). For instance, small working teams in math and engineering dominated by men reduce women's belonging, performance, and interest in pursuing STEM careers, and also activate gender stereotypes about STEM and with it, anxiety (Dasgupta et al., 2015). Women are evaluated more negatively in mixed-gender teams, where their performance is underestimated by male teammates (Grunspan et al., 2016). Increasing the percentage of women teammates produces a corresponding increase in women's team identification and collective efficacy (Niler et al., 2020). Men's sense of team identification, collective efficacy, and team performance is not influenced by gender composition. Parallel effects of gender composition on organizational performance and group decision-making have been well documented using both observational and experimental data; effects persist even after various controls (Apesteguia et al., 2012; Chen & Houser, 2019).

Collaboration is fundamentally about communication and coordination among people whose thoughts, feelings, and actions influence others' thoughts, feelings, and reactions. Team problem-solving interactions involve socio-cognitive processes such as active participation, conversational grounding, perspective taking, self and coregulation, behavioral mirroring, and monitoring of joint actions, all of which are externalized in communication that is dynamic, emergent, and measurable (Dowell et al., 2019; Dowell & Poquet, 2021). The following section outlines two opportunities and four recommendations afforded by generative AI in addressing these disparities and fostering inclusive STEM team environments.

**Leveraging Generative AI for Inclusion in STEM Teams**

How might AI help dismantle the barriers for underrepresented students in STEM and foster their persistence in the pipeline? This section presents two opportunities: AI in collaborative assessments, and AI in adaptive, personalized systems to support individuals and STEM teams in collaborative problem-solving. Four policy recommendations follow: 1) formalize collaborative skills assessment; 2) increase inclusivity in analytical approach with fine-grained modeling; 3) enhance support for socio-cognitive and emotional learning research; 4) leverage GenAI in Human-AI teaming to promote inclusive collaboration.

**Assess Skills and Model Learning Processes for Inclusivity**

Collaborative problem-solving is considered a core 21st-century skill (Graesser et al., 2018; OECD, 2017). Successful collaboration requires strong social and emotional competencies. The key to a welcoming and inclusive STEM climate includes building skills in emotional understanding and management, self-awareness, and social perspectives, for all team members to experience greater connection and collective intelligence (Schneider et al., 2021).

Assessing collaborative skills and processes can be challenging (Graesser et al., 2018). While collaborative skills are yet to be fully integrated into formal assessments in higher education institutions, students' success in certain STEM courses relies heavily on team projects and the quality of teamwork. The conventional assessment of students' collaboration in teams is self-reporting, intra-group peer assessment (Gueldenzoph & May, 2002). This may invoke potential subjectivity bias and stereotypes toward minoritized group members or elicit concerns about unfair assessment (Sherrard, Raafat & Weaver, 1994) and further perpetuate inequality. For instance, women are evaluated more negatively in mixed-gender teams (Heilman & Haynes, 2005), where their performance is underestimated by male teammates (Grunspan et al., 2016). Furthermore, members of groups with more women, regardless of gender, received more negative judgments of individuals' task contributions, the group's effectiveness, and the desire to work with one's group again in a course of traditionally male-dominated subjects (West et al., 2012). However, some evidence suggests online self and peer assessments of group contribution, claims that women receive higher ratings than men (Tucker, 2014). As a result, the equitable assessment of collaborative learning challenges diverse learners.

To reduce biases often found in self-reported evaluations, an objective evaluation strategy focuses on analyzing learners' natural interactions (Dowell & Kovanovic, 2022; Schneider et al., 2021). This technique offers a more precise and impartial assessment of the dynamics of collaboration, thus enhancing the reliability of the evaluation process in collaborative settings. Teamwork skills and team dynamics appear through verbal and nonverbal behaviors; these behaviors can be captured and analyzed with machine learning and natural language processing (NLP) techniques (Chopade et al., 2019). Although large language models (LLMs) inherently come with concerns for biases, multimodal large language models can be grounded within learning theory and real-world practice to support equity and inclusion in collaborative learning (Lewis, 2022). Additionally, intervention to increase representation has focused on women in STEM, whereas other underserved groups received less attention (Sloane et al., 2021). To this end, NLP-driven methods allow diagnosing areas in which different subgroups of learners require support.

The use of AI approaches in education for collecting trace data, such as text and discourse, allows for the detection of teaching and learning patterns. This method provides intricate insights into learners' natural behaviors. Currently, researchers are exploring whether these behavioral patterns vary based on students' socio-demographic backgrounds and reveal areas where interventions could be most impactful in reducing disparities. This approach represents a significant step towards personalized and equitable education. For example, Dowell, Lin et al., (2019) modeled STEM collaborative interactions and examined how variations in team gender composition (female-minority, sex-parity, and female-majority) impact socio-cognitive conversation patterns among team members using an AI-based NLP technique, Group Communication Analysis. Their results revealed novel insights, suggesting that the behavioral impact of male-dominated teams was more specific than simply sex differences in speaking up. Instead, these differences are evident in the extent to which they engage in productive discourse that responds to what other learners said previously (*overall responsivity*), provide meaningful contributions that warrant follow-up by peers (*social impact*), and monitor and build on their own previous contributions over the course of interaction (*internal cohesion*). Women's conversation shows greater overall responsivity, social impact, and internal cohesion than that of men. Furthermore, they found that men engaged in similar productive discourse when more women were present in the group. These findings break the image of how women are typically hindered in groups and deliver an empowering message for women who are aspiring to enter male-dominated fields.

**Recommendation 1: Formalized Collaborative Skills Assessment**

　　　Formal assessment of collaborative socio-cognitive and emotional skills should be ubiquitous in STEM education. An emphasis on collaborative skills building is conducive to creating a more sustaining impact on equity in the future by preparing individuals to be more sensible and perceptive to individuals of different backgrounds in teams. Indeed, these measures are already being incorporated in large tech companies, where we find that in addition to technical competence, evaluators often assess applicants based on their ability to communicate effectively with a range of audiences and function effectively on a team (Hirudayaraj et al., 2021; Rios et al., 2020). However, there is currently little consensus on how collaboration or group work should be measured or evaluated. It is important that we develop standards for research and practice regarding the measurement of collaborative skills. Collaborative assessment should be pertinent to individuals' collaborative skills as well as optimizing group dynamics (Andrews-Todd & Forsyth, 2018). When developing new standards and protocols for assessments, we suggest a connected view that would provide clear rubrics for educational practitioners, as well as frameworks that are suitable for developing automatic assessments in technology-enhanced environments. Combining self or peer assessment, teacher observations, as well as AI assessments, we can achieve a more comprehensive view of collaborative learning that balances potential biases from solely subjective reports or algorithmic modeling across classrooms and computer-supported environments.

**Recommendation 2: Toward Inclusive Analytics**

AI's ability to capture fine-grained learner behavior can revolutionize collaboration evaluation, focusing on process-oriented assessments. Future research might explore how certain collaborative processes benefit learning experiences, particularly for underrepresented students, and identify group dynamics that enhance a sense of belonging. As Mercier et al. (2023) pointed out, many existing quantitative studies in collaborative learning fail to account for the intersection of multiple groups (e.g., Black + Females in STEM), when minoritized students are aggregated into one group. AI's capacity to detect nuanced learner behaviors enhances the potential of collaborative analytics to model diverse learners at scale, particularly in recognizing and examining intersectionalities. These efforts should be at the front agenda for future inclusive collaborative analytics. Additionally, this approach toward inclusive analytics calls for incorporating feedback from teachers and students to shape these new assessment methods. Ensuring transparency and explainability of AI in this context is crucial, necessitating the involvement of educators and learners in the development of equitable evaluation metrics.

**Adaptive and Personalized Support for Diverse Teams**

Collaborative learning environments, enhanced by AI, offer personalized experiences tailored to individual needs. Intelligent tutoring systems (ITS) employ advanced AI models to support students' learning paths, tracking progress, and enabling tutors to provide targeted guidance (Rovira et al., 2017). Dialogue-based ITSs use large language models for personalized, context-aware feedback (Grenander et al., 2021). In team learning, AI agents sensitive to psychological and problem states aid in collaborative problem-solving (Graesser et al., 2018). While effective, these systems often prioritize cognitive support over psychological aspects. Research suggests a need for integrating pedagogical and psychological theories in ITS development (Zawacki-Richter et al., 2019), and for incorporating inclusivity and equity in AIEd (Ouyang & Jiao, 2021).

In line with this, Generative AI has demonstrated the capacity to model conversations with teams of students. For instance, generative AI such as ChatGPT can provide human-machine conversation and increase adaptation and interactivation in digital environments. This highlights the opportunities stemming from the integration of AI in team collaboration, specifically the shift in perception of AI from being mere *tools* to acting as *teammates* (Seeber et al., 2020). This concept revolves around AI systems actively participating in team dynamics, decision-making, and problem-solving processes. Instead of just facilitating tasks, AI as teammates involves a deeper integration into the collaborative environment. Contributing to and influencing team interactions and outcomes. This approach emphasizing a more symbiotic relationship between AI and human team members, where AI's role transcends utility and becomes integral to the team's functioning, potentially leading to more dynamic and effective collaboration in STEM.

**Recommendation 3: Funding Socio-Cognitive and Emotion Research**

Incorporating adaptive, personalized support from GenAI can significantly enhance socio-cognitive and emotional learning (ScEL) in diverse STEM teams. Funding in this area would enable the development of GenAI tools that adaptively respond to the unique dynamics and needs of each team, fostering better communication, empathy, and collaboration. This personalized approach ensures that teams benefit from targeted interventions, promoting inclusivity and equity in STEM team environments (Dowell et al, 2020). Such funding would not only support the technological development of these advanced AI systems but also the necessary research, training, and policy development to integrate them effectively into STEM education and professional practice.

**Recommendation 4: Human-AI Teaming for Inclusive Collaboration**

GenAI can be a powerful tool in STEM education to address issues of unequal participation and microaggressions. AI-driven simulations of diverse role models in STEM could challenge stereotypes and inspire broader participation, as suggested by Cohen et al. (2019). More empirical studies are needed to explore the idea of leveraging GenAI to construct inclusive awareness training or curricula for students. This builds upon GenAI's capacity to simulate interactive collaborative scenarios and enact certain types of behaviors that disrupt or promote inclusivity in teams. In the foreseeable future where GenAI is integrated into STEM collaboration as teammates (Seeber et al., 2020), more personalized and adaptive learning experiences can cater to diverse team dynamics. GenAI teammates can help mitigate biases, support underrepresented group members, and promote equitable participation. Policies should promote the development of AI systems that understand diverse team dynamics, enhancing inclusive participation. Ethical AI development can ensure systems free from biases and respectful of diverse perspectives. Additionally, policies could advocate for training programs that equip STEM professionals with the skills to effectively collaborate with AI teammates, ensuring that these integrations enhance team performance while maintaining a focus on equity and inclusion.

**Call to Action: Collaborative Endeavor between Stakeholders**

AI offers opportunities that require researchers, educators, and policymakers to unite in their effort towards creating an equitable STEM ecosystem. GenAI such as chatGPT is often not developed with educational goals in mind. This requires careful consideration of context, data collection, and measurement validity when deploying AI into learning systems. When it comes to driving STEM diversity and inclusion, involving underrepresented students' voices in developing policies and actions around building any integrated AI systems is imperative. The top-down approach in conventional educational policy formation and implementation would not be effective if we do not involve critical opinions of the population such policies are serving. Considering divergent perspectives will also facilitate transparent discussions for trust building towards AI use. Lastly, we must also ensure expert opinions on fairness and ethics are consulted, scrutinize the potential algorithmic bias that AI could introduce, and how we should use certain technology in moderation to eliminate potential harmful consequences.

In advancing research and practice, a successful collaboration between researchers and educational technology developers can bring enormous impact. The business sector has a powerful reach to a broader mass of learners and offers platforms for collaborative educational technology to be tested. However, the balance between business motivation and research validation is a tricky one to strike, and policies are needed to ensure the latter is prioritized. While diversity and equity is an important mission both sides agree upon, businesses rarely prioritize it on the agenda due to the obscurity of measuring the impact and financial return. To address this, AI-driven assessments could offer quantifiable data insights that help promote actionable steps on the business agenda. Lastly, policies incentivizing strong collaboration between small-scale business innovation and educational research could help foster a future where there are more diverse technological incubations specifically aimed at promoting diversity and equity.

## Conclusion

The integration of Generative AI in STEM education presents a transformative opportunity to address the enduring challenges of diversity and inclusion. Harnessing GenAI for collaborative problem-solving and personalized support can mitigate psychological barriers faced by underrepresented groups, enhancing their participation and success. However, this requires a collaborative effort among educators, policymakers, and technology developers to ensure ethical and effective implementation. Policies should emphasize the formal assessment of collaborative skills, inclusive analytics, funding for socio-cognitive and emotional research, and the development of human-AI teaming strategies. These areas are not comprehensive, but instead merely the tip of the iceberg. For example, future work needs to be done surrounding intersectionality, accessibility including assistive technology for disability, and inclusion for neurodiversity (see also O'Brien et al., 2022). This all requires a continuous coordinated commitment from researchers, educators, policymakers, and technologists—who have the collective responsibility to harness the power of AI to create diverse, inclusive, and equitable STEM ecosystems that empower all learners to pursue their STEM aspirations. The path forward demands concerted action, innovative thinking, and a steadfast commitment to making STEM a field where diversity is not just welcomed but celebrated as a cornerstone of its strength and vitality.

**Harnessing Generative AI for Inclusion and Diversity** Nixon 17

**Harnessing Generative AI for Inclusion and Diversity**　　　　　　　　　　　　Nixon　　　19
/